**Title**

# Challenges and opportunities to computationally deconvolve heterogeneous tissue with varying cell sizes using single cell RNA-sequencing datasets

**Authors**

Sean K. Maden[1], Sang Ho Kwon[2,3], Louise A. Huuki-Myers[2], Leonardo Collado-Torres[2], Stephanie C. Hicks[1,4,*], Kristen R. Maynard[2,3,5,*]

**Affiliations**

1. Department of Biostatistics, Johns Hopkins Bloomberg School of Public Health, Baltimore, MD, USA
2. Lieber Institute for Brain Development, Johns Hopkins Medical Campus, Baltimore, MD, USA
3. The Solomon H. Snyder Department of Neuroscience, Johns Hopkins School of Medicine, Baltimore, MD, USA
4. Malone Center for Engineering in Healthcare, Johns Hopkins University, Baltimore, MD, USA
5. Department of Psychiatry and Behavioral Sciences, Johns Hopkins School of Medicine, Baltimore, MD, USA

*Equal contribution and co-corresponding senior authors

**Abstract**

Deconvolution of cell mixtures in “bulk” transcriptomic samples from homogenate human tissue is important for understanding the pathologies of diseases. However, several experimental and computational challenges remain in developing and implementing transcriptomics-based deconvolution approaches, especially those using a single cell/nuclei RNA-seq reference atlas, which are becoming rapidly available across many tissues. Notably, deconvolution algorithms are frequently developed using samples from tissues with similar cell sizes. However, brain tissue or immune cell populations have cell types with substantially different cell sizes, total mRNA expression, and transcriptional activity. When existing deconvolution approaches are applied to these tissues, these systematic differences in cell sizes and transcriptomic activity confound accurate cell proportion estimates and instead may quantify total mRNA content. Furthermore, there is a lack of standard reference atlases and computational approaches to facilitate integrative analyses, including not only bulk and single cell/nuclei RNA-seq data, but also new data modalities from spatial -omic or imaging approaches. New multi-assay datasets need to be collected with orthogonal data types generated from the same tissue block and the same individual, to serve as a “gold standard” for evaluating new and existing deconvolution methods. Below, we discuss these key challenges and how they can be addressed with the acquisition of new datasets and approaches to analysis.

# Introduction

An important challenge in the analysis of gene expression data from complex tissue homogenates measured with RNA-sequencing (bulk RNA-seq) is to reconcile cellular heterogeneity, or unique gene expression profiles of distinct cell types in the sample. A prime example is bulk RNA-seq data from human brain tissue, which consists of two major categories of cell types, neurons and glia, both of which have distinct morphologies, cell sizes, and functions across brain regions and sub-regions (1–3). Failing to account for biases driven by molecular and biological characteristics of distinct cell types can lead to inaccurate cell type proportion estimates from deconvolution of complex tissue such as brain (3).

Broadly, methods that computationally estimate cell proportions from bulk tissue “-omics” data, such as gene expression or DNA methylation (DNAm) data, are referred to as “deconvolution algorithms” (4,5). Deconvolution commonly uses three terms: (1) a cell type signatures reference atlas, called *Z*; (2) a convoluted signals matrix, *Y*; and (3) a vector of the proportions of cell types in *Y*, called *P*. Here, we focus on gene expression reference-based algorithms that predict *P* given *Z* and *Y* (**Figure 1**).

Recent work has described important challenges (**Figure 2**) for deconvolution with various tissues including blood, kidney, and pancreas (6,7). However, tissues with notably different cell sizes, total mRNA expression, and transcriptional activity levels, such as brain or immune cell populations, present additional challenges for deconvolution that have not yet been described in the literature. It is important to be able to accurately estimate the cell composition of these tissues, as the cell composition has been shown to change with disease (8–13).

In computational methods development, gold standard datasets are used to set baseline performance expectations and provide a well-characterized reference against which new outputs can be evaluated. For example, Sanger sequencing is used as a gold standard platform for validation of genetic sequencing data (14,15). In deconvolution, independent or orthogonal measurements (**Figure 3**) from different platforms of cell composition can be used to validate algorithm-based estimates from bulk tissue expression.

In this paper, we summarize a set of challenges for performing deconvolution in highly heterogeneous tissues, using human brain tissue as a motivating example. We also present a set of recommendations and future opportunities for how to address these challenges to more accurately estimate tissue cell composition and better understand human disease. This poses an opportunity to set a higher bar for biological discovery and publication practices including increased computational reproducibility (7). The ability to iteratively implement and optimize new methods and benchmark workflows in heterogeneous tissues will enable deconvolution tools to further our understanding of the role of changes in cell type composition with disease risk and progression.

# Challenge 1: Lack of orthogonal measurements to evaluate deconvolution results across samples, donors, platforms, and studies

**Need for orthogonal measurements from matched tissue samples for bulk and single cell data.** When developing a deconvolution method, using matched bulk and single cell/nucleus RNA-seq (sc/snRNA-seq) datasets from the same tissue samples (**Figure 3**) enables controlling for potential confounding of biological variation, specifically donor-to-donor variation (16,17). Biological variation can be an important confound for deconvolution experiments. For example, Wang et al., 2019 (16) studied errors from using a sc/snRNA-seq reference dataset from source A to deconvolute a RNA-seq sample from source B can lead to inaccurate estimates of cell composition for source B, where sources could be distinct donors or studies.

**Need for orthogonal measurements from health and disease samples.** Deconvolution algorithms are commonly used to investigate whether changes in cell composition of tissue samples are associated with a phenotype or outcome, such as in case-control study designs. This poses a potential generalizability challenge when algorithms (**Table 4**) are only trained on one type of tissue sample (e.g. healthy/control samples) and not on tissues with the observed phenotype or outcome (e.g. disease samples). It was previously shown (18) that differential expression (DE) between group conditions can limit the utility of a normal tissue reference to accurately deconvolve cell type abundances in a disease condition. Including multiple phenotypes can also avoid algorithm overfitting, encourage selection of better cell type markers, and boost the overall generalizability of findings. Ideally, cases should be matched to the reference samples on potentially confounding factors like subject demographics, tissue collection procedures, and specimen handling strategies.

**Need for orthogonal measurements to form a reference atlas (Z) across multiple donors.** A key experimental design consideration is to select the sc/snRNA-seq samples used to build a reference atlas (Z). For example, a reference atlas (Z) could contain data from multiple donors or from only tissue samples that have matched bulk and sc/snRNA-seq samples. This decision depends on the specific research question, the statistical power to detect cell types (19), availability of previously published data (5), and the cost of generating new data (20). Multi-group references can mitigate the low reliability of cell type proportion estimates from a single sc/snRNA-seq sample (18). As sc/snRNA-seq data is characteristically sparse, pooling cells across groups can further boost power to characterize rare, small, or less active cell types (19,21).

**Need for measurements of cell type composition from orthogonal platforms.** The primary gold standard measurement to evaluate the accuracy of estimated cell compositions from a deconvolution algorithm is an orthogonal cell type fraction measurement (**Table 1**) in the tissue sample, and this should ideally be known with high accuracy and reliability. In multiple tissues including blood and brain, fluorescence-activated cell sorted (FACS) RNA-seq (22,23) and DNAm microarray data (3,24) have been used as orthogonal measurements of "true" cell composition. Cell type proportion estimates based on relative yields from sc/snRNA-seq data are not likely to be reliable (22) because of dissociation bias (25) and incomplete representation of sequenced cells (i.e. only a subset of the sample is sequenced). This bias impacts the "true" cell composition yield in a cell type-specific manner (26), is not present in bulk RNA-seq data, and can explain systematic expression differences between bulk RNA-seq data (27). As a solution, orthogonal cell type measures could ideally be extracted from many different data types (**Table 1**), including microscopy images from molecular marker-based protocols such as single molecule fluorescent in situ hybridization (smFISH) (3). This allows for characterization of cell type proportions as well as other size/shape measurements directly from the tissue. Emerging spatial transcriptomics technologies further integrate gene expression with precise coordinates from image data (28). While platforms such as Visium (10x Genomics) (29) yield spatial transcriptomics data at 55μm "spot" resolution containing multiple cells, technologies such as MERFISH (30) and Xenium (31) generate data at single cell resolution (32).

## Challenge 2: Cell types vary in abundance, size, and total mRNA

**Cell types exhibit a wide range in size and function within and across human tissues**. Most eukaryotic cells are between 10-100μm in diameter, for example ranging from red blood cells (8μm), skin cells (30μm), and neurons (up to 1m long) (33). In particular, the brain is an excellent example of a tissue exhibiting a wide range of cell types with different sizes and morphologies (7,34). Within the brain, there are a diversity of cell types that fall into several broad categories, including neurons, glia, and vasculature-related cells. These cell types have distinct functions reflected by differences in morphology, physiology, cell body size, and molecular identity. For example, neurons are larger and more transcriptionally active than glial cells (2). Vasculature-related cells, including endothelial cells, smooth muscle cells, and pericytes that comprise the building blocks of blood vessels and are also smaller in size than neurons (35). These cell types have specific genetic programs that facilitate distinct functions (35). For example, neurons (larger excitatory glutamatergic neurons and smaller inhibitory GABAergic neurons (36)) are larger and less numerous than glial cells, a heterogeneous group of cells comprised of oligodendrocytes (Oligo) (20-200μm) (37), oligodendrocyte precursor cells (OPC) (50μm) (38), microglia (15-30μm) (39), and astrocytes (Astro) (40-60μm) (40), which serve many roles, such as myelination, immune signaling, and physical and metabolic support. This extensive cell type diversity found in the brain, and other tissues, underscores the motivation for adjusting for differences in cell sizes prior to performing deconvolution (see data sources in **Table 1**).

**Cell type scale factor transformations can improve the performance of deconvolution algorithms.** While bulk transcriptomics deconvolution commonly predicts cell type proportions from expression data, it was noted that this approach may instead quantify total mRNA content in the absence of an adjustment for systematic differences in size and expression activity at the cell type level (3). This adjustment, which we will call a 'cell type scale factor transformation' (or cell scale factors for short), is used to transform the reference atlas (Z) data prior to deconvolution (3,41). It was introduced for microarray-based expression data (41,42) and later used for scRNA-seq data in multiple tissues (3,43,44). Cell scale factors are frequently used to generate sc/snRNA-seq-based data that resemble real bulk RNA-seq data based on "pseudobulking" or aggregating molecular profiles across sc/snRNA-seq data (45). Reference atlas transformation using orthogonal and non-orthogonal cell scale factors reduced errors from deconvolution-based cell proportion predictions. This may be because estimates without this transformation quantify total RNA rather than cell proportion (3). Cell scale factors may be estimated from either expression or expression-orthogonal data, such as sorted or purified populations of immune cells, which are used in existing deconvolution algorithms such as *EPIC* and *ABIS* (43,44). The algorithms *MuSiC* and *MuSiC2* (16,18) can use either expression-based or user-defined scale factors (**Table 4**). Importantly, there are currently no standards for applying cell scale factors prior to deconvolution, and users may need to transform the reference atlas (Z) prior to calling certain algorithms. Further, many algorithms have not been extensively tested in complex tissues, such as brain, that show large differences in size and transcriptomic activity across cell types. Ultimately, more reliable cell scale factor estimation and standardized transformation procedures can facilitate future deconvolution research (3,41).

**Different approaches to obtain cell scale factors can influence cell composition estimates.** There are several approaches to estimate and scale cell types in application of deconvolution. Expression-orthogonal cell size estimation methods can come from, for example, fluorescent in situ hybridization (FISH) or immunohistochemistry (IHC) (3,27,36) (**Table 2**). Image processing softwares such as ImageJ/FIJI (46) and HALO (Indica Labs) can provide cell body or nucleus measurements, including diameter, area, perimeter, among other size features (**Table 3**). However, cell segmentation presents a key obstacle limiting the accuracy of imaging-based approaches, especially for cells with complex morphologies (47). Expression-based cell size estimates are commonly calculated from total mRNA counts, often referred to as "library size factor" (48), which are typically unique to each cell, but could also be considered distinct for each cell type (**Table 4**). However, these estimates may be confounded by either the total sequenced RNA or genes with outlying high expression (43). For this reason, total expressed genes may be a good alternative robust to this type of confound. Cell scale factors from sc/snRNA-seq data are further subject to bias from tissue dissociation, cell compartment isolation, and other factors that have cell type-specific impacts (16–18). Another consideration is the application of cell scale factor transformations, as published deconvolution algorithms apply scale factors before (16) or after (41) prediction of cell type proportions. Application of cell scale factor transformation to the reference atlas (Z) may prevent quantification of total RNA rather than cell proportions (3). In summary, cell scale factor transformations can improve bulk transcriptomics deconvolution across multiple species, tissues, and sequencing platforms.

## Challenge 3: Protocol bias for tissue processing impacts reference atlas (Z)

**Acquisition of data with single nucleus (sn) versus single cell (sc) RNA-seq protocols.** A reference atlas (Z) from individual cells may be obtained from the whole cell or just the nuclear compartment, which has been demonstrated as representative of the whole cell (49,50). In the human brain, the majority of studies are conducted on fresh frozen post-mortem tissue rather than fresh tissue. When post-mortem brain tissues are flash frozen during the preservation process, cells are lysed prohibiting the molecular profiling of whole single cells using scRNA-seq approaches. Instead, only nuclei are accessible for profiling using snRNA-seq approaches. While the nuclear transcriptome is representative of the whole cell transcriptome (51–53) nuclear transcripts include more intron-containing pre-mature mRNA and may not include transcripts locally expressed in cytoplasmic compartments, such as neuronal axons and dendrites, or transcripts rapidly exported out of the nucleus (2). On the other hand, compared to whole cells, nuclei are less sensitive to mechanical/enzymatic tissue dissociation procedures, which may artificially impact gene expression (25), and are suitable for multi-omic profiling such as combined RNA-seq and ATAC-seq from the same nucleus (54). In fact, dissociation protocol differences help explain the large differences in average nuclei per donor observed across brain snRNA-seq reference datasets (10). Importantly, reference datasets from the human brain (**Figure 4**) are often restricted to nuclear information while bulk RNA-seq brain data contains both nuclear and cytoplasmic information. While prior work showed only a small impact from cell compartment DE between bulk and snRNA-seq data, accounting for this slightly improves deconvolution accuracy (55). However, new computational methods are being developed to remove these protocol-specific biases (16).

**Tissue preparation protocols can impact the diversity and quality of cells profiled during sc/snRNA-seq.** Cell type-specific associations between dissociation treatment and gene expression were observed from sc/snRNA-seq data across multiple tissues and species (25). Expression patterns may further be influenced by the specific cell/nucleus isolation protocol utilized (25,56). There are several approaches for isolating nuclei from frozen tissues and removing debris from homogenization steps. While some studies employ a centrifugation-based approach with gradients of sucrose or iodixanol to purify nuclei from debris (57,58), others use fluorescence-activated nuclear sorting (FANS) to label and mechanically isolate single nuclei (59,60). FANS also allows for enrichment of distinct cell types by implementing an immunolabeling procedure for populations of interest prior to sorting. There are pros and cons to each of these nuclei preparation approaches. FANS gating strategies may bias towards certain cell sizes and influence the final population of profiled cells. In the brain, recent work highlighted advantages for sorting approaches that remove non-nuclear ambient RNA contaminating glial cell populations (61). Ultimately, tissue dissociation protocols can drive variation among and between sc/snRNA-seq populations.

**Choice of sc/snRNA-seq platforms can impact reference gene expression profiles.** There are several sequencing platform technologies to generate sc/snRNA-seq reference profiles. While these have been previously reviewed (20,62), it is important to note that the different sample preparations and chemistries required for each of these platforms impacts the downstream gene expression data. For example, the widely used single cell gene expression platform from 10x Genomics is a droplet-based approach offering a 3' or 5' assay for up to 10,000 nuclei/cells in a single pooled reaction (63). While the 10x Genomics platform allows profiling a large number of cells in a single experiment, a major limitation is the sparsity of data and restriction of coverage to one end of the transcript. This is in contrast to approaches such as SMART-seq (64) from Takara, which offers full-length transcriptome analysis, but requires isolation of nuclei into individual tubes for separate reactions, thereby often resulting in fewer total cells profiled. Other technologies are rapidly becoming available for sc/snRNA-seq approaches, and each of these can introduce different biases into reference data. Importantly, recently published deconvolution algorithms use data transformation strategies to adjust for these biases (16,27).

**Potential differences in library preparation strategies for bulk RNA-seq and sc/snRNA-seq data.** Library preparation is a crucial protocol step impacting RNA profiles in RNA-seq data. The two most popular strategies are ribosomal RNA (rRNA) depletion (65,66), where rRNA is removed and remaining RNA sequenced, and polyA-enrichment (67), where polyA mRNA is isolated and sequenced. The former strategy can isolate a more diverse RNA population, including pre-mature and alternatively spliced mRNAs lacking polyA tails, and non-protein encoding RNAs (68,69). This difference may drive protocol bias that needs to be accounted for (70). Library preparation strategies may differ between bulk and sc/snRNA-seq data used for deconvolution. While polyA-enrichment was initially common for bulk RNA-seq, many newly available datasets now use rRNA depletion. By contrast, with the accessibility and popularity of the sc/sn droplet-based technologies (63), many reference atlases (Z) are based on polyA-enrichment. Further, marker genes may not be consistently expressed across different library preparation conditions, which can reduce deconvolution accuracy. As newer deconvolution algorithms accept large marker gene sets, systematic RNA population differences between library preparation strategies likely need to be accounted for, warranting further investigation.

**Assay-specific biases between bulk and sc/snRNA-seq data.** Systematic differences between bulk RNA-seq and sc/snRNA-seq assays can increase errors and reduce the utility of estimated cell type abundances from deconvolution algorithms. These biases may arise from differences in sample processing protocol (e.g. cDNA synthesis, PCR amplification, UMI versus full-length transcript, etc.), sequencing platform (e.g. short- versus long-read, droplet- or microfluidics-based, etc.), and cell compartment isolation (e.g. whole cell, only cytoplasm, or only nucleus) (71,72). Different sequencing technologies also show varying transcript length bias, which increases power to detect highly expressed long transcripts over low expressed short transcripts (73,74). This bias can impact the genes and pathways identified from DE analyses (75,76). While use of unique molecular identifiers (UMIs) protocols (74,77) may reduce the extent of transcript length bias in sc/snRNA-seq data relative to bulk, it may persist from internal priming, a type of off-target polyA primer binding (78). Furthermore, unlike bulk RNA-seq datasets, sc/snRNA-seq data are highly sensitive to both cDNA synthesis and PCR protocols (71). Great improvements to both protocols have been made in recent years (79,80). Finally, bulk and sc/snRNA-seq data show distinct distributional properties that may impact downstream analyses and the utility of simulation approaches (45,81). Dispersion, or the extent of inequality between expression variances and means, is among the most important of these (82). Bulk RNA-seq expression may show less dispersion, and thus may be modeled either using a Poisson or negative binomial (83) distribution, while expression sparsity and heterogeneity in sc/snRNA-seq data increases dispersion and often motivates use of the negative binomial distribution (84,85).

**Differences in detectability of rare cell types across batches and assays.** Because cell type detection from sc/snRNA-seq data is confounded by low expression levels, downsampling sc/snRNA-seq profiles on library size is often performed prior to downstream analyses (86). Recently introduced normalization strategies can further increase the reliability of rare cell type quantification (16), and similar approaches are already being applied to newer spatial sc/snRNA-seq datasets (87). This may be especially useful for complex heterogeneous tissues like brain, where previously noted protocol biases limit the amount of available reference data for rare cell types (7). In general, uncommon or rare cell types do not have a large impact on abundant cell type predictions unless there is high expression collinearity between gene markers of rare and abundant cell types (6). In the human brain, deconvolution accuracy decreased substantially with the exclusion of neurons, but not less common glial cell types (55). Importantly, the low-end limit for reliable cell type proportion predictions was found to vary across deconvolution algorithms (88).

## Challenge 4: Standardization of cell type annotation and marker selection strategies

**Standard brain cell type definitions and nomenclature are complex and emerging.** As new cell type-specific molecular and functional datasets rapidly come online, our understanding and definition of cell type diversity is evolving. In the context of the brain, key factors impacting our understanding of distinct cell populations (89) include 1) discovery and improved molecular characterization of functionally distinct cell types in brain regions and subregions, 2) new insights into how physiology and connectivity impact neuronal identity, and 3) an improved understanding of how cells change during development and aging. Anatomical and spatial position also influences cell type gene expression. For example, while virtually all excitatory populations in the cortex are glutamatergic pyramidal neurons, they show strong molecular and morphological differences across cortical layers (90) and still further differences with glutamatergic populations in other brain regions such as the hippocampus and amygdala (59). This underscores the necessity for a common cell type nomenclature to organize cell type labels and pair these with key contextual features like tissue microenvironment (89). Further, as new data emerge and cell type nomenclature evolves, reference datasets will likely need to be revisited and modified accordingly to ensure their utility.

**Cell type resolution should be experimentally driven.** Given that cell type definitions can be complex and defined at multiple resolutions (i.e. as either broad cell classes or as fine subpopulations), the resolution for a given deconvolution analysis needs to be experimentally motivated. That is, the ideal cell type resolution may differ depending on the biological question under investigation. For certain applications, such as distinguishing the contribution of two adjacent brain regions to a given bulk RNA-seq sample, relatively coarse definitions of neurons and glial cells may be adequate. For other applications, such as understanding the contribution of different neuronal cell types to differential gene expression between healthy and disease samples, fine resolution cell types may be required. An important first step for deconvolution is deciding the appropriate cell type resolution to address the underlying biological question. Prior work in human blood utilized an optimization procedure to identify the 17 most optimal blood and immune cell types for deconvolution from 29 total candidate cell types (43). In the human brain, it was found that definition of the reference atlas (Z) is more important than the choice of deconvolution algorithm, and accordingly the target cell types should have expression data of sufficient quality to select the most optimal marker genes possible (55).

**Cell type definitions should be based on robust and identifiable expression data.** One of the key conditions of a successful deconvolution experiment is that the cell types of interest are identifiable in the sample type(s) of interest. For a cell type to be identifiable, it should be sufficiently abundant and have clear gene markers. Gene markers should have sufficient expression to be distinguishable from background (i.e. relative high expression and sufficient read depth), as well as from other cell types of interest (i.e. sufficient DE from other cell types, with other cell types ideally having none or very low expression) (88). While reference-free deconvolution algorithms (91–93) do not rely on specific reference marker genes to the same degree as reference-based algorithms, the suitability of available expression data to perform deconvolution with high accuracy is a key issue across algorithm types and needs to be carefully considered.

Even with appropriate cell type definitions and evidence from expression data, the issue of defining the total cell types (K) to predict in a sample presents its own challenge. If the cell types in the reference do not reflect the cell types in the bulk or pseudobulk sample, deconvolution accuracy can suffer (6). Given a set of more than two well defined cell type labels, it is also reasonable to ask whether we should deconvolve all cell types together, or whether similar cell types should be binned prior to attempting deconvolution. For example, suppose an expression dataset contains cells with the Excit, Inhib, Oligo, and Astro cell type labels. From these, we could define the following K=4 types, each with its own reference atlas: (1) neuronal (i.e. excitatory and inhibitory) and non-neuronal (i.e. Oligo and Astro); (2) Excit, Inhib, and non-neuronal; or (3) Excit, Inhib, Oligo, and Astro. Recent deconvolution studies have advanced our understanding of how cell type label definitions impact deconvolution outcomes. In both blood (43) and brain (55), iterative assessments may lead to the effective quantification of relatively specific cell types and exclusion of others. Efforts to bin and evaluate cell type definitions should be considered alongside strategies to identify the cell type-specific gene markers for the reference. Marker identification methods may be based on differences in differentially expressed genes, such as Wilcoxon rank sum statistics, and clustering, to name a few (94).

**Expression markers of disease may confound signature atlas reliability.** A further consideration for bulk deconvolution methods is heterogeneity introduced by disease state that may influence marker gene expression. As many algorithms are intended for use in bulk tissue samples from disease states, it is important to understand how illness may uniquely impact cell types and their expression of core marker genes. For example, in samples from individuals with Alzheimer's Disease (AD) relative to neurotypical control subjects, neurons show marker gene repression, while glial cells generally show up-regulation of marker genes (9). Changes in gene expression have also been reported for psychiatric disorders such as major depression, where prior work showed 16 cell types with altered expression including excitatory and OPC cell types (8). Given that disease-specific differential expression can interfere with the effectiveness of cell type signature matrices, cell type marker genes selected for deconvolution should show equivalent expression between healthy and disease conditions. If expression is not equivalent between conditions, further adjustments to either the reference marker or bulk expression data may be necessary.

## Challenge 5: Reference atlases (Z) should be built on standardized and state-of-the-art computational tools and file formats

**Standardized data-driven cell type labels can facilitate deconvolution advances.** As discussed above, effective cell type definitions are crucial for deconvolution success. As more data comes online (**Figure 4**), there is increasing need for uniform labeling of cell types (7) and careful documentation of study metadata, including cell type enrichment methods (95,96). For example, in the brain, anti-NeuN antibodies are commonly used to enrich neuronal cell populations during FANS (97). Cataloging cell markers and the reagents used to select specific cell types will be important for standardizing data collection practices. On the data analysis side, sc/snRNA-seq cell type labels may be derived from clustering (43,59,98), reference-based tools (99,100), or other analytical approaches (88,101,102). In these cases, cell type labels could be indexed with a link to their originating annotation method. Further, hierarchical organization of cell type descriptors can facilitate insights into their molecular and physiological properties. Examples of this practice include term ontologies from the ENCODE project (https://www.encodeproject.org) and CCN (89), and it can be leveraged for cell type marker selection (102). In summary, combining key analysis and definitional metadata with standardized cell type labels can encourage reproducibility and new analyses.

**Expression data needs to be published using state-of-the-art data science formats.** Publishing key datasets and analytical results with essential documentation and using standard data formats is an important part of reproducible computational research (103–106). While flat table files (e.g. files with .csv or .tsv extension) are most common, many other data formats allow rapid and memory-efficient access. Some important examples include relational database formats (e.g. structured query language [SQL], hierarchical data format 5 [HDF5]). These data formats are compatible with increasingly used cloud servers and remote computing environments (107). Further specialized data formats include the *SummarizedExperiment* format for most -omics data types (108), and the *SingleCellExperiment* format for sc/snRNA-seq expression data (48,109), which is being extended for use with image coordinate information from spatial transcriptomics experiments (34,90,110,111). Newer data formats may be subject to updates that introduce errors or conflicts with other data classes, and resolving data class conflicts frequently demands a high degree of technical knowledge. This is one reason it is important to publish versions along with packages and object classes, in case an older version needs to be used while a newer version is updated. In summary, sequencing data may be published in a variety of formats to facilitate access, and methods should include details like versions for computational tools that were used.

## Challenge 6: Improving algorithm and signature atlas generalizability to new bulk tissue conditions

**Cross-validation can limit algorithm overfitting and improve algorithm generalizability.** Developers of new deconvolution algorithms and studies seeking to benchmark existing approaches must consider statistical power (112) and generalizability (113). Here, power refers to the ability to detect cell type markers from DE analysis and differentiate between significantly different cell type proportions (19) and generalizability refers to the replicability of the experiment (103,114). For example, an experiment showing good algorithm performance in terms of accurate cell composition estimates and reliable cross-group comparisons could also perform well when analyzing additional data from an independent data source or new participant population. To encourage generalizability and reduce chances of algorithm overfitting to training data, cross-validation should be performed whenever possible, even if sc/snRNA-seq reference data is only available from relatively few sources (114,115). As mentioned previously, subjects and sample characteristics should further be balanced across experimental groups, as imbalances could bias the results or undercut their generalizability (11).

**Developers should account for the tissues and conditions in which new algorithms will be applied** Deconvolution algorithms have varying performance across tissues and conditions, which we will call "domains", and algorithms may be considered either general (e.g. good performance across domains) or domain-specific (e.g. good performance in a specific domain). Further, algorithm assumptions may vary depending on their intended domains of use. For example, algorithms often assume good markers are known for each type when developed with normal tissues (88) but algorithms for bulk tumor deconvolution may assume no tumor cell type markers are available (41,44,116). As algorithms are often developed in a single or constrained domain set (**Table 4**) and then benchmarked in new domains, certain programming practices can facilitate algorithm testing across domains. For example, functions for algorithms like EPIC (44) and MuSiC (16,41) flexibly support either default or user-specified cell scale factors, which may encourage more standard application of these adjustments in deconvolution experiments. Ultimately, developers should carefully consider the scope and nature of the domain(s) in which an algorithm will be applied.

**Deconvolution algorithms should be optimized for prediction across conditions of interest.** Beyond understanding normal tissue expression dynamics, effective deconvolution can allow new hypothesis-testing to elucidate relationships between cell types and disease mechanisms. Of particular interest in brain research is the prospect of studying significant changes in the abundances of neurons and/or glial cells between neurotypical samples and neurodevelopmental, neuropsychiatric, and neurodegenerative disorders, including autism spectrum disorder (ASD), Parkinson's disease (PD), and AD. Glia-specific inflammation in AD is detectable from snRNA-seq data, and further studies could reveal biomarker candidates and risk factors with utility for patient prognosis or diagnosis (12). Microglial activation has been correlated with AD severity, illuminating mechanisms related to disease progression (27). Total neuron proportion may decline in AD brains and reflect neuronal death as a hallmark symptom of AD; this trend was detectable in bulk tissue using multiple deconvolution methods (27). Finally, accurate cell type quantification in case/control studies of bulk tissues revealed 29 novel differentially expressed genes in ASD that were independent of cell composition differences (55). As new data and algorithms are published, more practical guidelines (22,88) will be needed to match the most appropriate strategies to their specific biological questions.

## Future opportunities and recommendations

We wish to highlight several opportunities for bulk transcriptomics deconvolution in heterogeneous tissues, including human brain. First, new reference datasets featuring multiple orthogonal assays from matched samples have huge potential to shape and inform new studies. Second, aggregation of published data into centralized repositories using standard data formats paired with structured and comprehensive metadata will increase the impact of new reference datasets and reproducibility of analyses based on these reference data. Finally, mitigating biases and improving statistical rigor in sample collection, experimental design, and training new deconvolution methods should greatly improve the efficacy of new deconvolution algorithms and benchmarking of existing and emerging algorithms. Applying a transformation reference atlas (Z) matrix using cell scale factors, such as in **Table 3**, may reduce errors in deconvolution predictions due to improved quantification of cell proportions rather than RNA amounts (3).

Researchers can take several steps to act on these opportunities. First, even studies with a small number of donors can improve their rigor by running technical replicates (i.e.multiple runs of the same assay) and biological replicates (i.e. multiple distinct samples or tissue blocks from the same donor). Further, deconvolution algorithms can be deployed as high-quality open-access software packages and made available in centralized curated repositories such as CRAN or Bioconductor (108). Finally, new research efforts can utilize existing references to perform validation and inform collection of new samples.

## Conclusions

While the rapidly evolving future of transcriptomics is promising, it will be important to not only address existing experimental and computational challenges in this field, but also anticipate future challenges. We have drawn on our collective research experience to detail the key challenges of designing experiments with technical and biological replicates, effective use and integration of different assays run on the same specimen or tissue block, performance of data analyses to improve statistical rigor and generalizability of findings, and publication of datasets with comprehensive and structured metadata and methods with runnable and versioned code. Taking proactive steps to address these challenges will facilitate studies of increasing scale and complexity while encouraging greater reproducibility.

## Data Availability

Code and data tables to reproduce panels in Figures 1 and 4 are available on GitHub (https://github.com/LieberInstitute/deconvo_commentary-paper).

## Abbreviations

- Single nucleus RNA-sequencing (snRNA-seq)
- Single cell RNA-sequencing (scRNA-seq)
- DNA methylation (DNAm)
- Dorsolateral prefrontal cortex (DLPFC)
- Alzheimer’s Disease (AD)
- Parkinson’s Disease (PD)
- Differential expression (DE)
- Fluorescence-activated nuclear sorting (FANS)
- Fluorescence-activated cell sorting (FACS)
- Common cell type nomenclature (CCN)
- Fluorescent in situ hybridization (FISH)

- Single molecule FISH (smFISH)
- Immunohistochemical (IHC)

# Declarations

**Ethics approval and consent to participate**
Not applicable.

**Consent for publication**
Not applicable.

**Competing interests**
The authors declare that they have no competing interests.

**Funding**
This project was supported by the Lieber Institute for Brain Development, and National Institutes of Health grant R01 MH123183. All funding bodies had no role in the design of the study and collection, analysis, and interpretation of data and in writing the manuscript.

**Author contributions**
SKM, KRM, and SCH wrote the initial draft and edited the manuscript. SHK, SKM prepared the figures. SKM prepared the tables. LCT and LAHM contributed to the conceptualization of the manuscript and provided comments on the draft. All authors approved the final manuscript.

**Acknowledgements**
We would like to thank Kelsey Montgomery, Sophia Cinquemani, and Keri Martinowich for the discussions and feedback of this manuscript. While an Investigator at LIBD, Andrew E. Jaffe helped secure funding for this work. Schematic illustrations were generated using Biorender.

# Figure legends

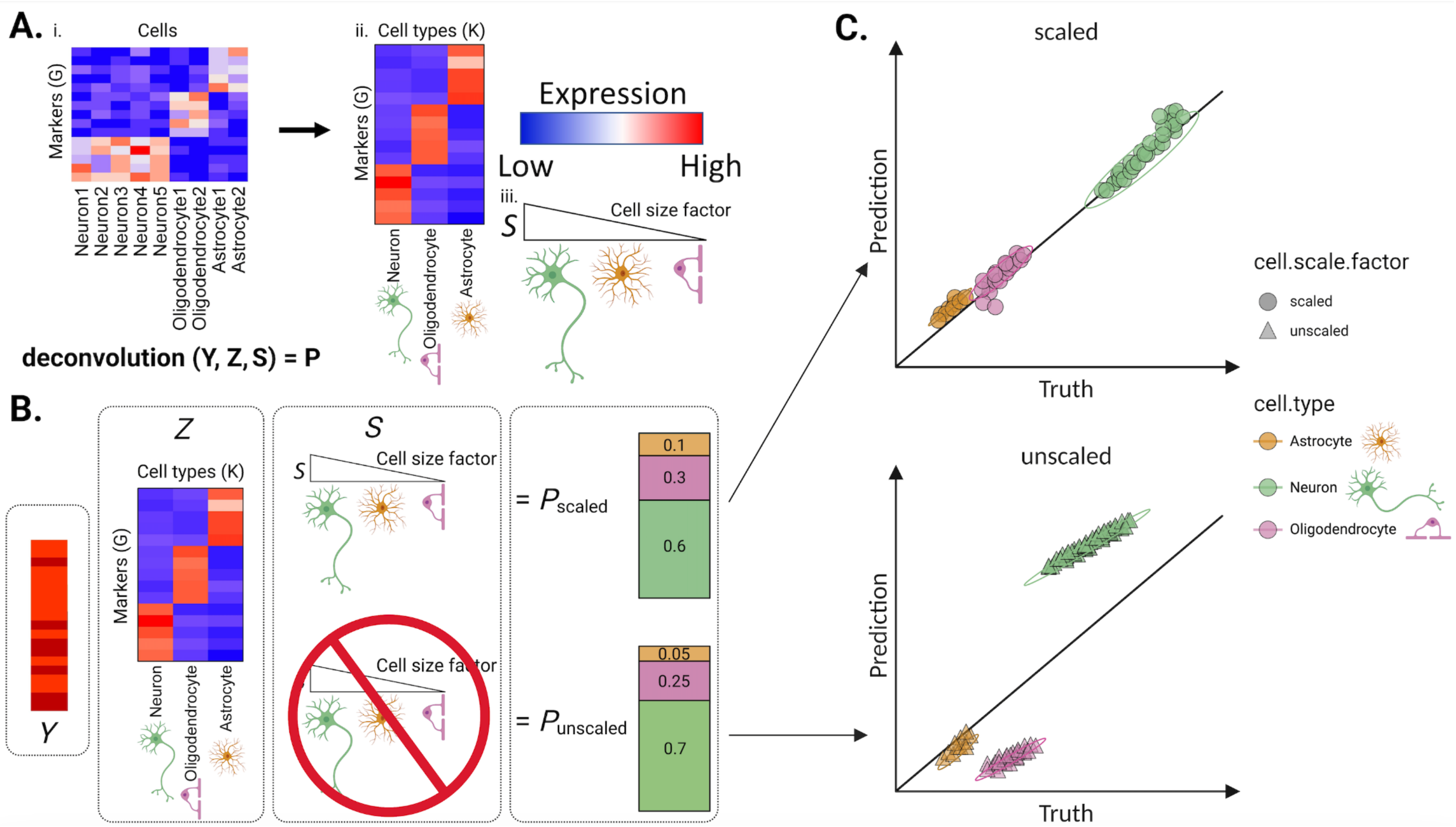

**Figure 1. Diagram of example deconvolution experiment using cell scale factors. A.** Heatmaps of gene expression: (**i**) for the (y-axis) marker genes G by cell labels for each of (x-axis) neurons, oligodendrocytes, or astrocytes, (**ii**) the (y-axis) G marker genes by (x-axis) cell types (*K*). Expression value colors: blue = low, white = intermediate, red = high. **(iii)** Wedge diagram of (*S*) cell scale factors, where wedge size is the value and cartoons indicate each cell type. **B.** (left-to-right) Heatmaps of bulk expression *Y*, and marker expression Z, cell scale factors *S*, and cell type proportions *P* for either (top) scaled or (bottom) unscaled expression, where bar plot values show cell type proportions with colors as in panel C. **C.** Scatterplot of example experiment results for multiple bulk samples *Y*, showing the (x-axis) true cell proportions and (y-axis) predicted cell proportions, where points are outcomes for a sample and cell type, and shapes show whether the cell scale factor transformation was applied. Plots were created using the ggplot2 v3.4.1 (117) and ComplexHeatmap v2.12.1 (118) software; data used to reproduce these plots are available from GitHub (Data Availability).

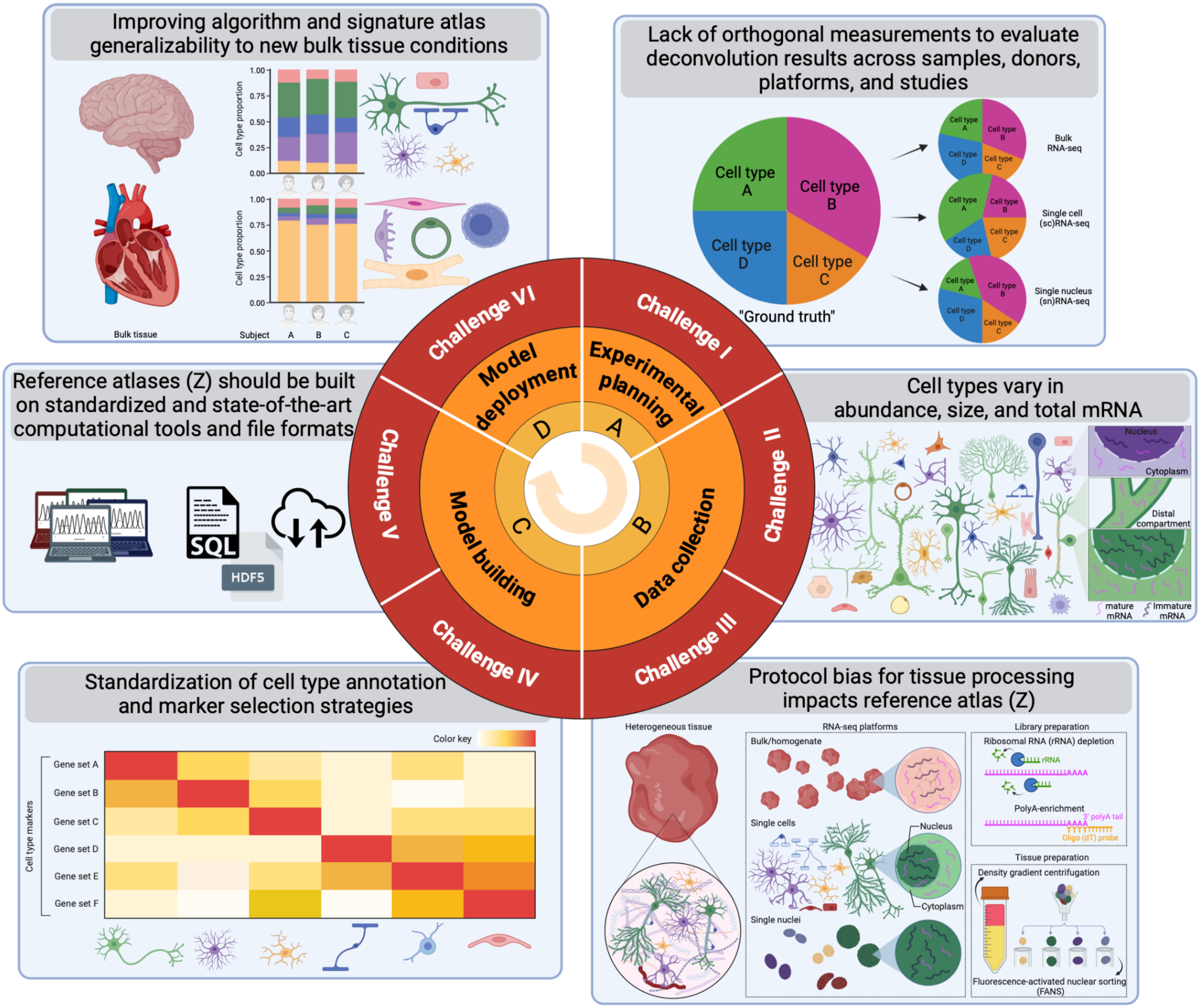

**Figure 2. Six challenges and opportunities to computationally deconvolve heterogeneous tissue with varying cell sizes using single cell RNA-sequencing datasets.** Direction of experimental process (middle arrow), experiment phases (orange labels), challenge number (red labels), challenge titles (gray panel titles), and depictions of key challenge concepts (box graphics).

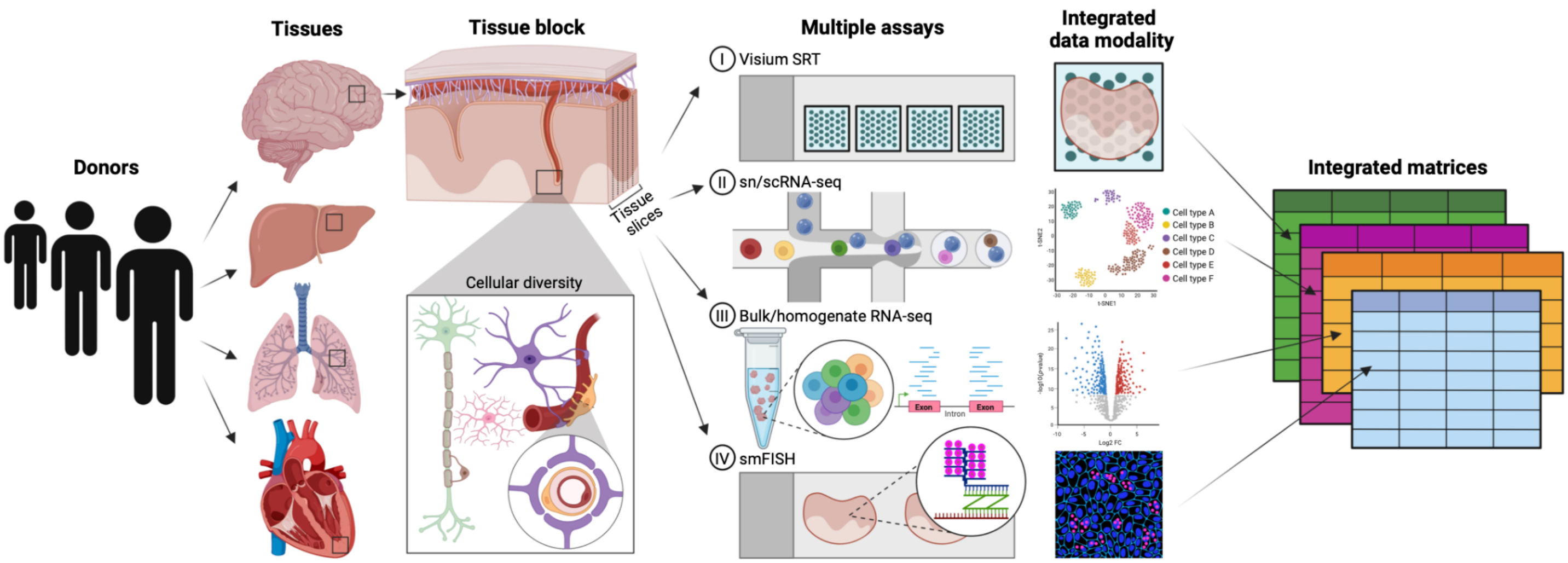

**Figure 3. Collecting an integrated dataset of orthogonal assays from the same tissue block across donors and tissues.** The development and benchmarking of deconvolution algorithms can be improved with gold standard reference datasets. Gold standards are developed across donors and tissues on which multiple assays are performed on the same tissue block. For example, adjacent sections of a tissue block could be used for spatial transcriptomics, sc/snRNA-seq, bulk/homogenate RNA-seq, and single molecule FISH (smFISH) to generate orthogonal cell type proportion and transcriptomic profile measurements. These assays generate data with distinct features (i.e. gene expression, cell size/shape, isoform diversity, etc) that can also be incorporated into deconvolution models to improve accuracy.

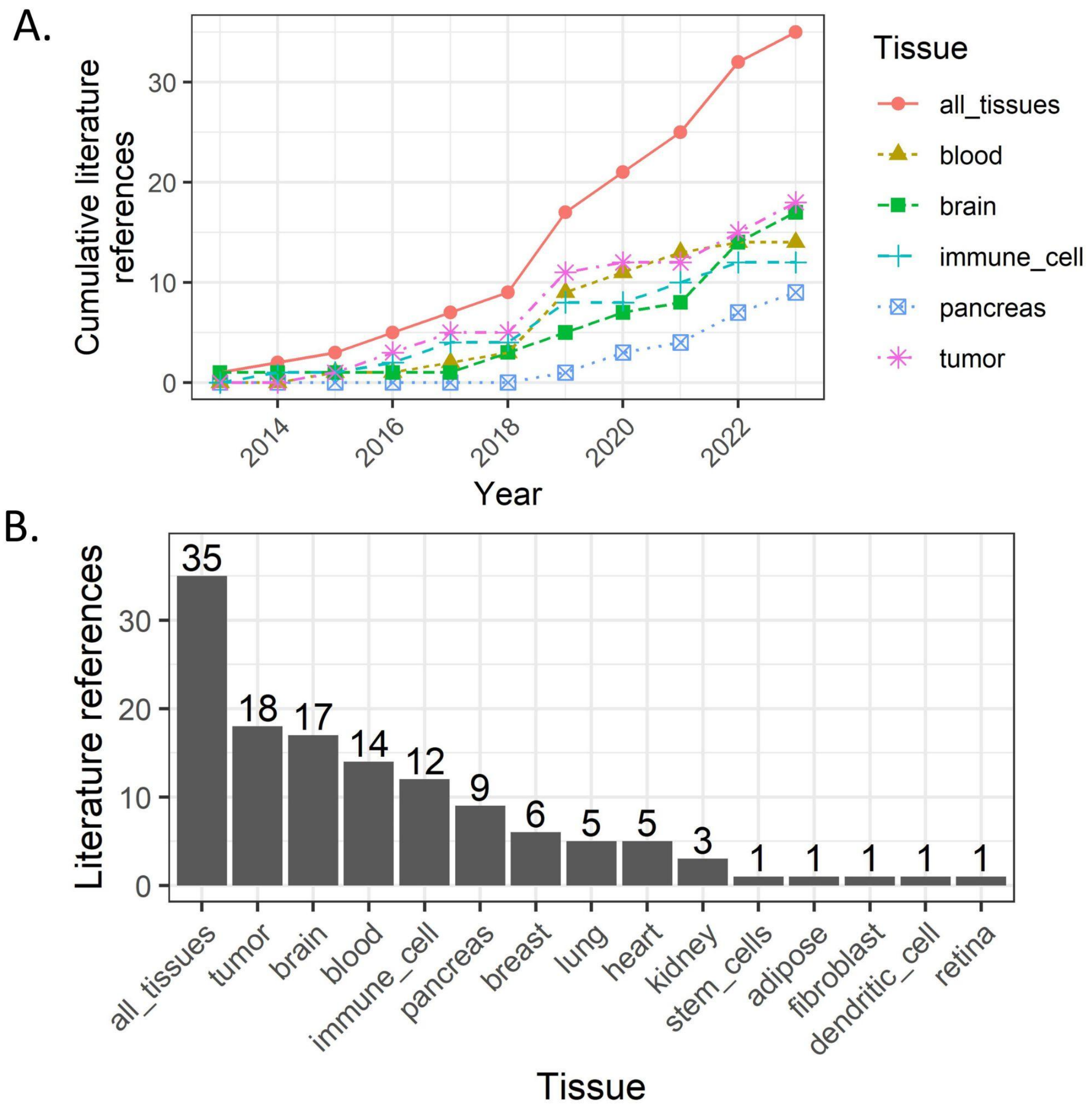

**Figure 4. Summary of tissues by literature reference from bulk transcriptomics deconvolution literature. A.** Dot and line plot of (x-axis) yearly (y-axis) cumulative references by (color, shape, line type) tissue, including (red, solid line, circles, "all_tissues" label) the combined set of all tissues. **B.** Barplot showing (y-axis) the number of literature references (x-axis) per tissue, including ("all_tissues" label) the combined set of all tissues. Plots were created using the ggplot2 (v3.4.1; (117) software; data used to reproduce these plots are available from GitHub (Data Availability).

# Table legends

**Table 1. Orthogonal cell type amount measurements used for bulk transcriptomics deconvolution.** Table describes the name (column 1) and a description (column 2) of the type of measurement, the type of assay used to capture the measurement (column 3), and example citations for these measurements (column 4).

| Name | Description | Assays | Citations |
|---|---|---|---|
| Fluorescent in situ hybridization (FISH) | Labeling and imaging of DNA-based cell type markers | In situ labeling, imaging | (3,119) |
| Immunohistochemistry (IHC) | Antibody-based cell marker labeling and imaging | In situ labeling, imaging | (44,120) |
| Immunofluorescence (IF) | Antibody-based fluorescent labeling of cell markers | In situ labeling, imaging | (3,121) |
| In vitro cell mixtures | Sequencing of manually mixed cells from dissociated bulk tissues or cell lines | Bulk RNA-seq | (17,55,122–125) |
| Fluorescence-activated cell sorting (FACS) | Sequencing of cells isolated by cytometric sorting | Flow cytometry; bulk RNA-seq | (22,43,44) |
| Genetic panel | DNA marker-based differentiation of tissues, esp. tumor from non-tumor | genetic marker assay; microarray | (124,126) |
| DNA methylation | Deconvolution using DNA methylation cell type markers | microarray; bisulfite sequencing | (3,24,127–130) |
| Hematoxylin and eosin staining | Clinical tissue slide staining procedure | In situ staining; imaging | (119,127) |

**Table 2. Experimental data platforms to estimate cell sizes and calculate cell size scaling factors to adjust for systematic differences in size and transcriptomic activity between cell types.** The table contains the type of experimental data (column 1), the metric used for cell size (column 2), a set of standards (gold, silver, and bronze) introduced by Dietrich et al. (2022) (column 3), the format for how the data are captured (column 4), example data analysis challenges when using these data (column 5), and if the experimental data are orthogonal to using sc/snRNA-seq (column 6).

| Experimental data | Cell size metric | Standard (45) | Data format | Data analysis challenges | Orthogonal to sc/snRNA-seq |
|---|---|---|---|---|---|
| FISH (4,101,131,132) | Label intensity | gold | Image | Label performance; cell segmentation; image artifact removal (16,18,43–45) | yes |
| IHQ/IHC (116) | Label intensity | gold | | | yes |
| Labeled expression marker (131,132) | Expression/label intensity | silver | | | yes |
| sc/snRNA-seq | mRNA spike-in expression | silver | Gene expression counts | Embedding alignment, batch effects, dissociation biases, platform biases (21,25,62) | yes |
| sc/snRNA-seq | Housekeeping gene expression | silver | | | no |
| sc/snRNA-seq | Library size (101,116,133) | bronze | | | no |
| sc/snRNA-seq | Expressed genes (101,116,133) | bronze | | | no |

**Table 3. Cell scale factor estimates from the literature, with focus on deconvolution studies that use sequencing references.** Values for blood cell types are from the SimBu R package (v1.2.0), and values for brain cell types are from Table 1 in (3). The Scale factor value (column 3) can be used in existing deconvolution algorithms leading to less biased results for estimating cell composition.

| Cell type | Tissue | Scale factor value | Scale factor type | Scale factor data source | Citation(s) |
|---|---|---|---|---|---|
| glial | brain | 91 | cell area | osmFISH | (3,132) |
| neuron | brain | 123 | cell area | osmFISH | (3,132) |
| glial | brain | 180 | nuclear mRNA | osmFISH | (3,132) |
| neuron | brain | 198 | nuclear mRNA | osmFISH | (3,132) |
| glial | brain | 12879 | library size | expression | (1,3) |
| neuron | brain | 18924 | library size | expression | (1,3) |
| B cells | multiple | 65.66 | median expression | Housekeeping gene expression | (45,116) |
| Macrophages | multiple | 138.12 | median expression | Housekeeping gene expression | (45,116) |
| Macrophages (M2) | multiple | 119.35 | median expression | Housekeeping gene expression | (45,116) |
| Monocytes | multiple | 130.65 | median expression | Housekeeping gene expression | (45,116) |
| Neutrophils | multiple | 27.74 | median expression | Housekeeping gene expression | (45,116) |
| NK cells | multiple | 117.72 | median expression | Housekeeping gene expression | (45,116) |
| T cells CD4 | multiple | 63.87 | median expression | Housekeeping gene expression | (45,116) |
| T cells CD8 | multiple | 70.26 | median expression | Housekeeping gene expression | (45,116) |
| T regulatory cells | multiple | 72.55 | median expression | Housekeeping gene expression | (45,116) |
| Dendritic cells | multiple | 140.76 | median expression | Housekeeping gene expression | (45,116) |
| T cells | multiple | 68.89 | median expression | Housekeeping gene expression | (45,116) |
| B cells | multiple | 0.40 | intensity | FACS | (44,45) |
| Macrophages | multiple | 1.42 | intensity | FACS | (44,45) |
| Monocytes | multiple | 1.42 | intensity | FACS | (44,45) |
| Neutrophils | multiple | 0.13 | intensity | FACS | (44,45) |
| NK cells | multiple | 0.44 | intensity | FACS | (44,45) |
| T cells | multiple | 0.40 | intensity | FACS | (44,45) |
| T cells CD4 | multiple | 0.40 | intensity | FACS | (44,45) |
| T cells CD8 | multiple | 0.40 | intensity | FACS | (44,45) |
| T helper cells | multiple | 0.40 | intensity | FACS | (44,45) |
| T regulatory cells | multiple | 0.40 | intensity | FACS | (44,45) |
| B cells | multiple | 20837.57 | intensity | FACS | (43,45) |
| Monocytes | multiple | 22824.32 | intensity | FACS | (43,45) |
| Neutrophils | multiple | 9546.74 | intensity | FACS | (43,45) |
| NK cells | multiple | 21456.91 | intensity | FACS | (43,45) |
| T cells CD4 | multiple | 14262.07 | intensity | FACS | (43,45) |
| T cells CD8 | multiple | 10660.95 | intensity | FACS | (43,45) |
| Plasma cells | multiple | 325800.99 | intensity | FACS | (43,45) |
| Dendritic cells | multiple | 57322.18 | intensity | FACS | (43,45) |

**Table 4. Deconvolution algorithms developed for bulk transcriptomics with sc/snRNA-seq reference datasets.** The table includes the name and reference (column 1) along with the year published (column 2) and a description (column 3) of the algorithm. The primary tissues used in the publication associated with the algorithm is also provided (column 4).

| Algorithm | Year | Description | Primary publication tissues |
|---|---|---|---|
| Coex (55) | 2022 | Marker co-expression networks and network module attribution | brain |
| MuSiC2 (18) | 2021 | Differential marker weighting and filtering on condition-specific differential expression | pancreas and retina |
| SCDC (17) | 2021 | Ensemble framework to integrate references across sources | pancreas and mammary gland |
| Bisque (27) | 2020 | Gene-specific transformations to address assay-specific biases | adipose and brain |
| MuSiC (16) | 2019 | Differential marker weighting to address marker expression confounding | pancreas and kidney |
| dtangle (134) | 2019 | Marker selection with linear mixed modeling | blood, breast, brain, liver, lung, muscle, cancer |
| ABIS (43) | 2019 | Absolute deconvolution with cell scale factors on TPM-normalized marker expression | blood and immune cells |
| quanTIseq (116) | 2019 | Non-negative regression with cell factor scaling and unknown cell type estimation | blood and tumor |
| Fardeep (135) | 2019 | Machine learning with adaptive trimmed least squares | tumor cells (GSM269529), immune cells (136) |
| BrainInABlender (100) | 2018 | Prediction with mean marker expression across references | brain, pyramidal neurons, stem cells, immune cells, blood cells |
| xCell (124) | 2017 | Linear scaling of marker enrichment scores | immune, stem, epithelial, and tumor cells |
| EPIC (44) | 2017 | Renormalization of reference markers by cell scale factors, quantification of unknown types | cancer and blood |
| MCP-counter (137) | 2016 | Cell type amount scoring for heterogeneous tissues, numerous cell types, and multiple clinical conditions | immune, stromal, and tumor cells and cell lines |
| TIMER (127) | 2016 | Batch effects removal form tumor purity markers; constrained least squares with orthogonal validation | multiple tumor types |
| CIBERSORT (138) | 2015 | Machine learning-based dimension reduction and permutation optimization | blood |
| DCQ (123) | 2014 | Whole transcriptome regularized regression followed by ensemble selection, with focus on cell surface marker genes | lung and immune cells |
| DeconRNASeq (122) | 2013 | Linear modeling, non-negative least squares, and quadratic programming | brain, heart, skeletal muscle, lung and liver |